\documentclass[showpacs,showkeys,floatfix,onecolumn]{revtex4}
\usepackage[cp1251]{inputenc}
\usepackage[english]{babel}
\usepackage[dvipdf]{color}
\usepackage{bm}
\usepackage{amsmath}

\begin{document}

\title{Landau theory for helical nematic phases}

\author{E.I.Kats and V.V.Lebedev}

\affiliation{Landau Institute for Theoretical Physics, RAS, \\
142432, Chernogolovka, Moscow region, Russia, and \\
Moscow Institute of Physics and Technology, \\
141700, Dolgoprudny, Moscow region, Russia}

\begin{abstract}

We propose Landau phenomenology for the phase transition from the conventional nematic into the conical helical orientationally non-uniform structure recently identified in liquid crystals formed by ``banana''-shaped molecules. The mean field predictions are mostly in agreement with experimental data. Based on the analogy with de Gennes model, we argue that fluctuations of the order parameter turn the transition to the first order phase transition rather than continuous one predicted by the mean-field theory. This conclusion is in agreement with experimental observations. We discuss the new Goldstone mode to be observed in the low-temperature phase.

\end{abstract}


\date{\today}

\maketitle

In the realm of liquid crystals one of the most exciting and relatively recent result is discovery of a new type of equilibrium nematic structure, see the papers \cite{HS09,PN10,CD11,BK13,MD14}. Especially surprising is the fact that the observed new phases (termed twist-bend nematics, $N_{TB}$) exhibit helical (chiral) orientational ordering despite being formed from achiral molecules. The molecules of the substances possessing $N_{TB}$ phases have a specific ``banana''-like shape. For comparison, known more than a century conventional nematic liquid crystals ($N$) are formed from rod-like or disk-like molecules. There exist also chiral cholesteric phases locally equivalent to nematics but possessing simple (orthogonal) helical structures with pitches in a few $\mu m$ range. The cholesteric structure appears as a result of relatively weak molecular chirality (that is why it has a relatively large pitch), and the swirl direction of the spiral (left or right) is determined by the sign of the molecular chirality. Unlike this situation, the $N_{TB}$ nematics are formed as a result of spontaneous chirality breaking, they have nanoscale pitches.

Scanning the literature one can find a number of theoretical works devoted to the twist-bend nematics \cite{DO01,KA12,AC13,ML13,SD13,VC14,VI14}). Majority of the works, starting from the influential paper \cite{DO01}, discuss the question how modulated orientational structures can be formed in achiral systems. One can easily understand that the description of the twist-bend nematics in terms of an orientational elastic energy requires a pathological (not positively defined) Frank elastic energy. An analysis in the framework of such Frank energy can rationalize some experimental observations made for the $N_{TB}$ liquid crystals, e.g., anomalously large flexoelectric coefficients \cite{SD13}, or non-monotonous temperature dependence of the orientational elastic moduli \cite{KA12}. Note the paper \cite{DD75} (which almost gone unnoticed for liquid crystal community), where the negative twist elasticity yielding to the spontaneous chiral symmetry breaking, has been suggested based on the Van der Waals contribution into the Frank elastic moduli. We note also the work \cite{VI14}, where $N_{TB}$ phase elasticity with two director fields have been discussed within the positively defined conventional Frank energy. In the recent preprint \cite{SA14} the authors consider how flexoelectricity combined with spontaneous polar order (ferroelectricity) could stabilize conic spiral orientational ordering. However, under natural Landau theory assumptions the theory \cite{SA14} yields to strongly biaxial and polar features of the $N_{TB}$ phase, apparently not supported by experimental observations.

From our point of view, a description of the twist-bend nematics in terms of an orientational elastic energy needs a modification related to relatively short pitch of the helicoidal structure. In the case the Frank moduli for the short-scale component of the director field are different from those for the long-scale component of the director. Therefore the components should be treated in terms of different elastic energies. In the paper we realize the program keeping the notation $\bm n$ for the long-scale component of the director and introducing its short-scale component $\bm\varphi$. The components have to be orthogonal, $\bm n\cdot \bm \varphi=0$. Thus the vector $\bm\varphi$ has two independent components.

The quantity $\bm\varphi$ plays a role of the order parameter for the phase transition $N$--$N_{TB}$. To analyze the system behavior in the vicinity of the transition, one should introduce the Landau functional in terms of $\bm\varphi$ including all relevant terms. The vector character of $\varphi$ leads to absence of odd terms in the Landau functional. Therefore in the mean field approximation the $N$--$N_{TB}$ transition is a continuous (second order) phase transition. Roughly, experimental data \cite{PN10,CD11,MD14} and numeric simulations \cite{KA12,ML13} can be positively confronted with the mean-field theory predictions. However, certain experiments \cite{CD11,MD14} clearly indicate that the $N$--$N_{TB}$ transition is not a continuous one: there exists a two-phase coexistence region where hysteresis phenomena are observed. The experimental data, which we aware, suggest: (i) a relatively weak first order $N$--$N_{TB}$ phase transition with barely visible fluctuation effects from the $N_{TB}$ side \cite{CD11,MD14}; (ii) practically regular and smooth temperature dependence of Frank elastic moduli \cite{KA12,AC13}. To explain the features one has to go beyond the mean field approximation and analyze fluctuational effects.

Experimentally, in the $N_{TB}$ phase the director $\bm n+\bm\varphi$ has the helical conic structure in space. By other words, the short-scale component $\bm\varphi$ rotates around $\bm n$ at moving along the $\bm n$-direction. Therefore the absolute value of the vector $\bm\varphi$ gives the tilt angle for the conical spiral. Because the conical helical structure has a certain short pitch periodicity (we characterize it by the wave vector $q_0$), the order parameter $\bm\varphi$ is condensed at passing to the $N_{TB}$ phase at a finite wave vector $q_0$ (experimentally on the order of a few inverse molecular lengths). Thus the $N$--$N_{TB}$ phase transition is similar to weak crystallization phase transitions \cite{KLM} where the mass density modulation appears at finite wave vectors. Besides, the vector nature of the order parameter leads to some peculiarities. Say, odd terms are absent in the Landau functional and some additional terms should be introduced there in comparison with the theory of weak crystallization.

For the vector order parameter the Landau functional contains second-order and fourth-order terms. Taking into account the nature of the short-scale vector field $\bm\varphi$, one obtains the following functional
 \begin{eqnarray}
 \int dV \left\{\frac{a}{2} \bm\varphi^2
 +\frac{b_3}{8 q_0^2} \left[
 \left(n_i n_k \partial_i \partial_k +q_0^2\right) \bm\varphi \right]^2
 +\frac{b_1}{2} (\nabla \bm\varphi)^2
 \right. \nonumber \\ \left.
 +\frac{b_\perp}{2}\delta^\perp_{ij}
 \partial_i \bm \varphi \partial_j \bm \varphi
 +\frac{\lambda}{24} \varphi^4
 -\frac{\lambda_1}{16 q_0^2}
 \left(\epsilon_{ijk} \varphi_i \partial_j \varphi_k\right)^2
 \right\}, \quad
 \label{bana1}
 \end{eqnarray}
where $\delta^\perp_{ij}=\delta_{ij}-n_i n_j$. As usual, $a\propto T-T_c$, where $T_c$ is the mean field transition temperature. The quantities $b$ are analogs of the Frank moduli for the order parameter $\bm\varphi$. The free energy (\ref{bana1}) represents the minimal Landau model for the $N$--$N_{TB}$ phase transition, catching all observable features of the $N_{TB}$ phase.

We first neglect fluctuations of the long-scale director $\bm n$ and assume that it is a homogeneous field $\bm n_0$, determining a preferred direction, $\bm n_0=(0,0,1)$. Then the Landau functional (\ref{bana1}) can be represented in a more compact form by replacing the order parameter $\bm\varphi$ by its complex counterpart $\bm\psi$
 \begin{equation}
 \bm\varphi= 2\,\mathrm{Re}\
 \left[\bm\psi \exp(i q_0 z) \right].
 \label{bana3}
 \end{equation}
Unlike $\bm\varphi$ the complex field $\psi$ is long-scale. It is perpendicular to $\bm n_0$, $\bm\psi= (\psi_x,\psi_y,0)$. In terms of the field $\bm\psi$, the Landau functional (\ref{bana1}) is rewritten as
 \begin{eqnarray}
 {\cal F}_\psi=\int dV \left\{a |\bm\psi|^2
 +b_\perp |\partial_\perp \bm\psi|^2
 +b_3 |\partial_z \bm\psi|^2
 +b_1 |\nabla\cdot \bm\psi|^2
 \right. \nonumber \\ \left.
 +\frac{\lambda}{4} (\bm\psi \bm\psi^*)^2
 -\frac{\lambda_1}{4}
 \left[ (\bm\psi \bm\psi^*)^2
 -\bm\psi^2 (\bm\psi ^*)^2 \right]
 \right\} . \qquad
 \label{bana2}
 \end{eqnarray}
If $\lambda_1>0$ then below the phase transition (at $a<0$) minimization of the last term in Eq. (\ref{bana2}) gives $\psi_x =i \psi_y$ or $\psi_x =-i \psi_y$ (in both cases $\bm\psi^2=0$). That corresponds just to the observed conical helical structure since then
 \begin{equation}
 \varphi_x=2 |\psi_x|\cos(q_0z+\phi), \quad
 \varphi_y=\pm 2|\psi_x|\sin(q_0z+\phi),
 \label{cone}
 \end{equation}
where $\phi$ is the phase of $\psi_x$ and signs $\pm$ correspond to two possible rotation directions of the conical structure.

It is worth noting one additional soft (Goldstone) mode in the $N_{TB}$ phase related to long-scale variations of the phase $\phi$ in the expression (\ref{cone}). Since the bulk energy is independent of a homogeneous phase shift, the elastic energy related to variations of $\phi$ depends on its gradient
 \begin{equation}
 {\cal F}_\mathrm{el}=\int dV\ \left[
 \frac{B_\perp}{2} (\partial_\perp\phi)^2
 +\frac{B_\parallel}{2}(\partial_z\phi)^2 \right].
 \label{super}
 \end{equation}
The energy (\ref{super}) defined at scales larger than the correlation length is analogous to the energy of the superfluid component in a superfluid helium. Unlike the helium, the energy (\ref{super}) is anisotropic.

The mean-field (i.e., ignoring fluctuations) predictions following from minimizing the energy (\ref{bana2}) are standard. Namely, at $T < T_c$ (in the low-temperature phase) $|\psi_x|\propto \sqrt {T_c - T}$. The specific heat has the standard Landau jump at the transition point, and the correlation length diverges as $(T_c - T)^{-1/2}$. In the mean field approximation the moduli in Eq. (\ref{super}) can be obtained from the functional (\ref{bana2}), they are $B_\perp=2(2b_\perp+b_1) |\psi_x|^2$, $B_\parallel=4 b_3 |\psi_x|^2 $. Therefore both, $B_\perp$ and $B_\parallel$, are proportional to $T_c-T$.

To proceed further we pass to an analysis of fluctuations. In our case, fluctuations of both components of the director, $\bm n$ and $\bm\varphi$, has to be taken into account. The long-scale director will be written as $\bm n=\bm n_0+\delta\bm n$, where $\delta\bm n$ is a relatively weak deviation of $\bm n$ from its average value. Next, it is convenient for us to keep the field $\bm\psi$ as representing components of the order parameter $\bm\varphi$ perpendicular to $\bm n_0$. Then in the linear over $\delta\bm n$ approximation the constraint $\bm n \cdot \bm \varphi=0$ leads to
 \begin{equation}
 \bm\varphi_\perp= 2\,\mathrm{Re}\
 \left[\bm\psi e^{i q_0 z} \right], \
 \varphi_\parallel= -\delta \bm n\cdot
 2\,\mathrm{Re}\
 \left[\bm\psi e^{i q_0 z} \right].
 \label{bana13}
 \end{equation}
instead of Eq. (\ref{bana3}). Here the subscripts $\parallel$ and $\perp$ mark the order parameter components parallel and perpendicular to $\bm n_0$.

Substituting the expressions (\ref{bana13}) into the gradient terms of the Landau functional (\ref{bana1}) and expanding over $\delta\bm n$, we find in the main approximation the following interaction terms
 \begin{eqnarray}
 {\cal F}_{int1} = \int dV\
 b_\perp \left[ i q_0 \delta n_\alpha
 (\partial_\alpha \bm\psi \cdot \bm \psi^*
 \right. \nonumber \\ \left.
 -\bm \psi \cdot \partial_\alpha \bm \psi^*)
 +q_0^2 (\delta\bm n)^2 |\bm\psi|^2  \right],
 \label{bana14} \\
 {\cal F}_{int2} = \int dV\
 b_1 \left\{iq_0 \left[
 (\delta\bm n\cdot \bm \psi) (\nabla\cdot \bm\psi^*)
 \right. \right.\nonumber \\ \left. \left.
 - (\delta\bm n\cdot \bm \psi^*)
 (\nabla\cdot \bm\psi) \right]
 +q_0^2 (\delta\bm n\cdot \bm \psi)
 (\delta\bm n\cdot \bm \psi^*) \right\}.
 \label{bana15}
 \end{eqnarray}
Note that the terms (\ref{bana14},\ref{bana15}) can be obtained from the second order term (\ref{bana2}) by passing to the ``covariant'' derivative $\partial_i \psi\to (\partial_i +iq_0 \delta n_i)\psi$. This is a consequence of the rotational invariance of the system. The interaction terms have to be added to the Landau functional (\ref{bana2}) and to the Frank energy
 \begin{eqnarray}
 {\cal F}_\mathrm{Fr}=\int dV\
 \left\{ \frac{K_1}{2} (\nabla\cdot\delta\bm n)^2
 +\frac{K_2}{2}\left[\nabla_\perp \delta n_i
 \nabla_\perp \delta n_i
 \right. \right.\nonumber \\ \left. \left.
 -(\nabla\cdot\delta\bm n)^2\right]
 +\frac{K_3}{2} (\partial_z \delta\bm n)^2 \right\}.
 \label{Frank}
 \end{eqnarray}

The above contributions to the Landau functional (\ref{bana2},\ref{bana14},\ref{bana15},\ref{Frank}) constitute the complete set of relevant terms, that determine the fluctuation effects. The region of developed fluctuations (realized near the phase transition) can be analyzed, say, in the framework of the so-called $\epsilon$-expansion \cite{WK74} based on the renormalization group (RG) procedure. There is the common belief that the RG-flow draws the system towards a symmetric state, in our case realized at $b_1=0$. Then our model becomes almost identical to the de Gennes model \cite{GP} describing the $N$--$Sm A$ phase transition, apart from the number of components of the order parameter: in our case the order parameter is the two-component vector $\bm\psi$ whereas the order parameter is scalar in the de Gennes model. Fluctuation effects in the de Gennes model were analyzed in the framework of the RG-procedure first in \cite{HL74} and then (for an arbitrary number of components of the order parameter) in \cite{LC78}. It was stated in the work that the zero-charge fix point is stable only for very large number of components of the order parameter. Thus, one expects that the fluctuations destroy the zero-charge fix point that physically means converting the phase transition to the first order. 

Thus, we expect that fluctuations of the director turn the $N$--$Sm A$ phase transition to the first order class that is in agreement with experimental data. In the situation fluctuation effects can be observed only in a narrow vicinity of the phase transition and not to be strongly exhibited. Therefore the mean-field predictions can cover the majority of the temperature interval near the phase transition. This looks to be true for the experimental data \cite{PN10,CD11,KA12,ML13,MD14}. Especially important are X-ray small-angle diffraction studies \cite{BK13} directly manifesting short-scale periodicity. The temperature dependence of the diffraction peak width at $T < T_c$ found in \cite{BK13} corresponds to the mean-field prediction $\propto (T_c - T)^{1/2}$. The cusp in the quantity observed near the phase transition probably signals about fluctuation effects.

At the moment of writing this manuscript we are not aware about any observation of the additional Goldstone mode in the $N_{TB}$ phase related to long-scale variations of the cone phase. Although optical scattering methods could identify the mode, in practice, realization of such experiment is not a simple issue. First, it requires a very accurate selection of polarizations for the incident and scattered beams polarized to exclude presumably much larger scattering by conventional director modes. Second, since the optical wave vector is smaller than the inverse pinch period $q_0$, the only second order scattering (proportional to the square of the $N_{TB}$ order parameter fluctuations) contributes to the light scattering intensity. To the point, an external magnetic field which suppresses the conventional long-scale director fluctuations can be very useful for the observation of the mode.

This work was funded by the Russian Science Foundation via grant 14-12-00475. It is our pleasure to thank O.Lavrentovich for helpful communications and discussions.

\end{document}